\documentclass[conference]{IEEEtran}
\IEEEoverridecommandlockouts
\usepackage{makecell}
\usepackage[colorlinks,
            linkcolor=black,  
            anchorcolor=black,
            urlcolor=black,
            citecolor=black,  
            ]{hyperref}

\usepackage{cite}
\usepackage{threeparttable}
\usepackage{multirow}
\usepackage{amsmath,amssymb,amsfonts}
\usepackage{algorithmic}
\usepackage{graphicx}
\usepackage{textcomp}
\usepackage{xcolor}
\def\BibTeX{{\rm B\kern-.05em{\sc i\kern-.025em b}\kern-.08em
    T\kern-.1667em\lower.7ex\hbox{E}\kern-.125emX}}
\begin{document}

\title{Study of Lightweight Transformer Architectures for Single-Channel Speech Enhancement\\
}

\author{\IEEEauthorblockN{Haixin Zhao, Nilesh Madhu}
\IEEEauthorblockA{\textit{IDLab, Department of Electronics and Information Systems} \\
\textit{Ghent University - imec}\\
Ghent, Belgium \\
\{haixin.zhao, nilesh.madhu\}@ugent.be}
}

\maketitle

\begin{abstract}
In speech enhancement, achieving state-of-the-art (SotA) performance while adhering to the computational constraints on edge devices remains a formidable challenge. Networks integrating stacked temporal and spectral modelling effectively leverage improved architectures such as transformers; however, they inevitably incur substantial computational complexity and model expansion. Through systematic ablation analysis on transformer-based temporal and spectral modelling, we demonstrate that the architecture employing streamlined Frequency-Time-Frequency (FTF) stacked transformers efficiently learns global dependencies within causal context, while avoiding considerable computational demands. Utilising discriminators in training further improves learning efficacy and enhancement without introducing additional complexity during inference. The proposed lightweight, causal, transformer-based architecture with adversarial training (LCT-GAN) yields SoTA performance on instrumental metrics among contemporary lightweight models, but with far less overhead. Compared to DeepFilterNet2, the LCT-GAN only requires 6\% of the parameters, at similar complexity and performance. Against CCFNet+(Lite), LCT-GAN saves 9\% in parameters and 10\% in multiply-accumulate operations yet yielding improved performance. Further, the LCT-GAN even outperforms more complex, common baseline models on widely used test datasets.

\end{abstract}


\section{Introduction}

Deep learning-based methods have come to dominate the field of speech enhancement, surpassing statistical approaches \cite{wang2018supervised}. 
These methods are continually improving and evolving with the progress being made on the development of more powerful deep learning architectures and mechanisms, such as transformers and attention \cite{lu23e_interspeech, abdulatif2024cmgan}. 
However, such improvements come with substantial computational complexity, which is a challenge for practical applications on edge devices. This has led to research focused on the development of lightweight speech enhancement models in resource-constrained scenarios. 

\textbf{Prior work:} Due to their sequential modelling capabilities, recurrent neural network (RNN) layers, such as long short-term memory networks (LSTMs) and Gated Recurrent Units (GRUs) can be used to efficiently learn spectral and temporal {\em context} of speech \cite{luo2020dual, le21b_interspeech}. 
Incorporating such elements within speech enhancement models results in a significant boost in performance - making these components indispensable in state-of-the-art (SotA) architectures \cite{dang2023first, schroeter2022deepfilternet2, schroeter2023deepfilternet3}. To reduce their complexity and parameters, grouping strategies were developed and applied to this component, which is usually the main computational bottleneck in most lightweight enhancement architectures \cite{schroeter2022deepfilternet2, schroeter2023deepfilternet3, schroeter2022deepfilternet, rong2024gtcrn, yang2024fspen, braun2021towards}. 
Complexity and footprint reduction were also addressed by compressing models and replacing standard convolutions with depth-wise separable convolutions \cite{rong2024gtcrn}. Other approaches endeavoured to address this problem at the signal {\em representation} level e.g., by reducing the frequency resolution using fewer sub-bands or equivalent rectangular bandwidth (ERB) features \cite{rong2024gtcrn, schroeter2022deepfilternet, schroeter2022deepfilternet2, schroeter2023deepfilternet3}.
Still other approaches have endeavoured to significantly reduce the model size by employing the aforementioned strategies to the extreme \cite{rong2024gtcrn, yang2024fspen}. However, while this leads to ultra-low complexity, such optimisation is usually paired with an inevitable degradation of the enhancement performance. 

Instead of focusing only on network refinements, DeepFilterNet and its subsequent iterations~\cite{schroeter2022deepfilternet, schroeter2022deepfilternet2, schroeter2023deepfilternet3}, consisting of two sub-networks, incorporated domain knowledge from statistical approaches. The first stage estimates the envelope using ERB features, while the second stage of deep filtering further enhances the periodic components. 
DeepFilterNet2$\&$3 can be considered SotA among lightweight models in terms of enhancement performance, albeit still with a considerable number of parameters. 
A coarse and fine two-stage CCFNet(Lite) was proposed in ~\cite{dang2023first}, for further reducing the model footprint while maintaining comparable complexity as DeepFilterNet2.

\begin{figure*}
\centerline{\includegraphics[width=1\linewidth]{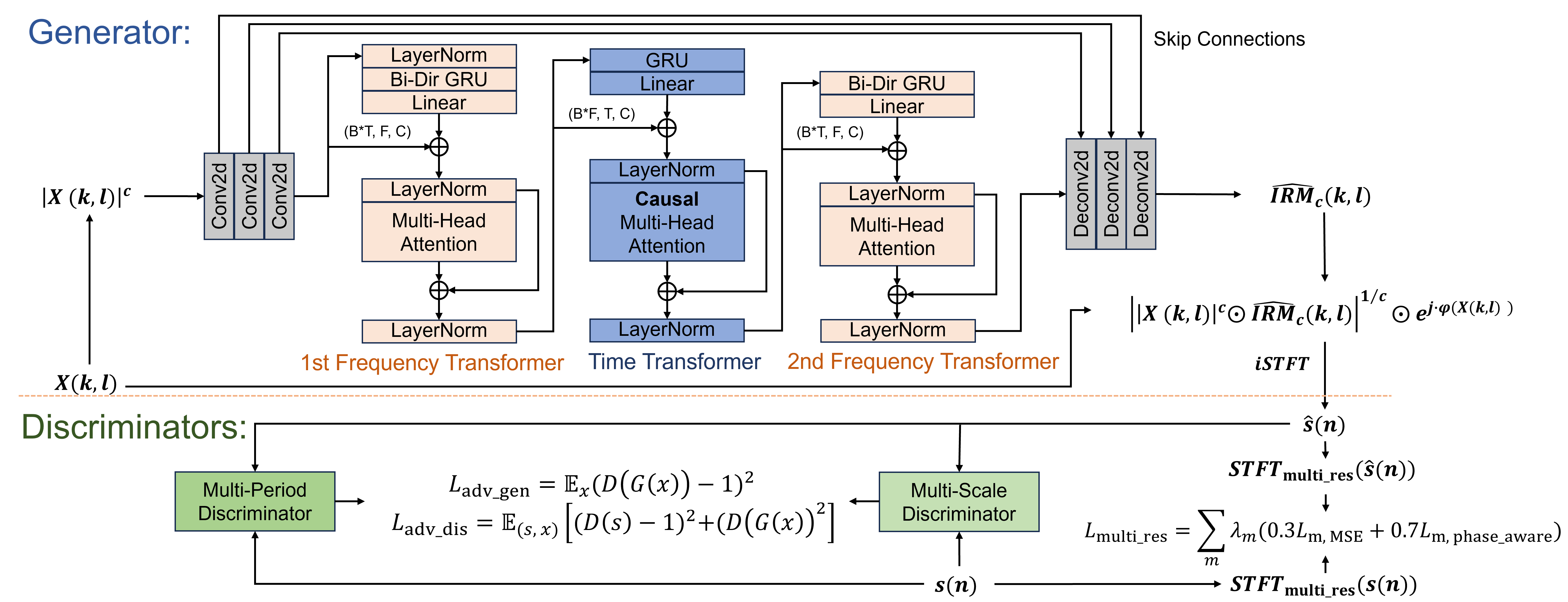}}
 \vspace{-0.28cm}
\caption{The architecture of the proposed LCT-GAN model, with the 'generator' and the discriminators distinguished by dotted lines. The generator is a predictive network to estimate masks for denoising.
The $L_\mathrm{adv\_gen}$ and $L_\mathrm{adv\_dis}$ denote the adversarial loss components for the training of the generator and discriminator, respectively. $G()$ denotes the generator (LCT), while $D()$ is the discriminator.
A multi-resolution loss, $L_\mathrm{multi\_res}$, is employed for the generator training. Skip connections are implemented by point-wise convolutional layers. The input tensor dimensions for each transformer are explicitly indicated. 
}  
\vspace{-0.45cm}
\label{fig1}
\end{figure*}

However, these reductions in parameters are concomitant with considerable performance decline, motivating the exploration of more efficient modelling architectures. Interleaving temporal and spectral modelling has demonstrated its impressive potential in enhancement performance \cite{lu23e_interspeech, luo2020dual}, yet the cumulative stacking of these modules results in a substantial increase in computational overhead, rendering it impractical for deployment on edge devices. 
Additionally, as more modules are stacked, the performance gains tend to become incremental, highlighting the necessity for investigation on efficient stacking strategies and the sequencing of temporal and spectral modelling.

\textbf{Contributions:} 
To surmount these challenges and gain deeper insights into the characteristics of Time-Frequency (TF) modelling, we conduct a systematic ablation study based on transformer blocks. Multi-Head Attention (MHA) layers are introduced to address the limitations of RNNs in capturing long-range dependencies and to dynamically attend to salient information, thereby contributing to modelling.

To further reduce the computational complexity and footprint, when modelling along one dimension (temporal or spectral), parameter sharing is applied across the other dimension. 
Consequently, the reduced feature size of GRUs and MHA blocks significantly scales down the model footprint, enabling the integration of superior yet computationally intensive attention mechanism in lightweight models.
However, such sharing prevents simultaneous dependency exploitation across the time and frequency dimensions. This is in contrast to existing methods which implicitly allow this by flattening the unmodeled dimension into the feature dimension~\cite{braun2021towards}.
Therefore, we build upon the insights from our ablation study, and further propose a streamlined Frequency-Time-Frequency (FTF) sequential modelling bottleneck, facilitating efficient global dependency exploitation across causal contexts. Trapezoidal masking is implemented in the time transformer to enforce causality while constraining computational overhead, to meet stricter latency requirements in certain scenarios.

Lastly, to further improve the preservation of speech components, without imposing additional overhead during inference, multi-scale and multi-period discriminators are employed in training. The corresponding ablation study shows the resulting benefits. With significantly fewer parameters and lower computational complexity, the proposed LCT-GAN achieves comparable (or better) enhancement performance to the above-mentioned SotA lightweight models and even outperforms more complex models.

\section{Method}

\subsection{Proposed LCT-GAN and Signal Model}
Fig.~\ref{fig1} illustrates the architecture of the proposed LCT-GAN model. The microphone signal $x(n)$, input to the network, is assumed to be an additive mixture of the target speech $s(n)$ and background noise, $v(n)$.
The LCT `generator' is a predictive model, estimating the underlying clean speech, $\widehat{s}(n)$, by applying a (real-valued) time-frequency mask to $X(k,l)$ (the short-time Fourier transform (STFT) domain representation of $x(n)$) followed by the inverse STFT. The generator training target is the compressed-domain ideal ratio mask (IRM), defined as:
\begin{equation}
\textstyle
    \widehat{IRM}_{\text{c}}(k,l) = \frac{\lvert S(k,l) \rvert^{c}}{\lvert X(k,l)\rvert^{c} + \gamma}
\end{equation}
where $k,l$ represent the frequency-bin and time-frames, and $\gamma$ is a regularisation constant. The compression factor ($c$) is set to 0.3 to refine the mask estimation for lower-energy speech components.
The mask in the linear domain is obtained by exponentiating $\widehat{IRM}_{\text{c}}(k,l)$.
Thus, we dedicate the network resources to learning solely a magnitude estimation, while retaining the distorted phase. This choice is ratified by previous work showing negligible gain from the prediction of complex-valued outputs \cite{wu23b_interspeech} and even degradation caused by compensation between magnitude and phase in these networks \cite{wang2021compensation}. 
We further validate this by benchmarking the proposed magnitude estimation against approaches with complex-valued output, all within the LCT framework.

The generator adopts a U-Net structure known for its high efficiency. Here, a 3-layer causal convolutional encoder-decoder is applied. Each encoder and decoder layer, except the output layer, is followed by a leaky ReLU with $\alpha=0.03$. Skip connections facilitate the transfer of latent features from the encoder to the decoder. 

To further improve the speech estimate, a network incorporating waveform-based multi-period and multi-scale discriminators is used, providing additional loss components through joint adversarial training. While the multi-period discriminator is utilised to capture long-term latent structures ~\cite{kumar2019melgan}, the multi-scale discriminator ~\cite{kong2020hifi} facilitates the efficient learning of periodic patterns.

\subsection{Efficient FTF Transformers}\label{sec.ftf}

Building upon the subsequent ablation study, we propose the FTF bottleneck with three lightweight sequential FTF transformers, effectively exploiting global information while significantly reducing the computation and parameter requirements in comparison to large-scale transformers in~\cite{lu23e_interspeech, abdulatif2024cmgan}.
Each transformer incorporates an RNN block and an attention block for global dependency exploitation. Residual connections are utilised, followed by Layer Normalisation for both the MHA blocks and GRUs. 
Trapezoidal causal masking is performed in the MHA block of the time transformer to maintain causality while limiting the context length to a maximum of 1 second. 
As a result, the complexity is independent of utterance length.

\begin{figure}
\centerline{\includegraphics[width=1\linewidth]{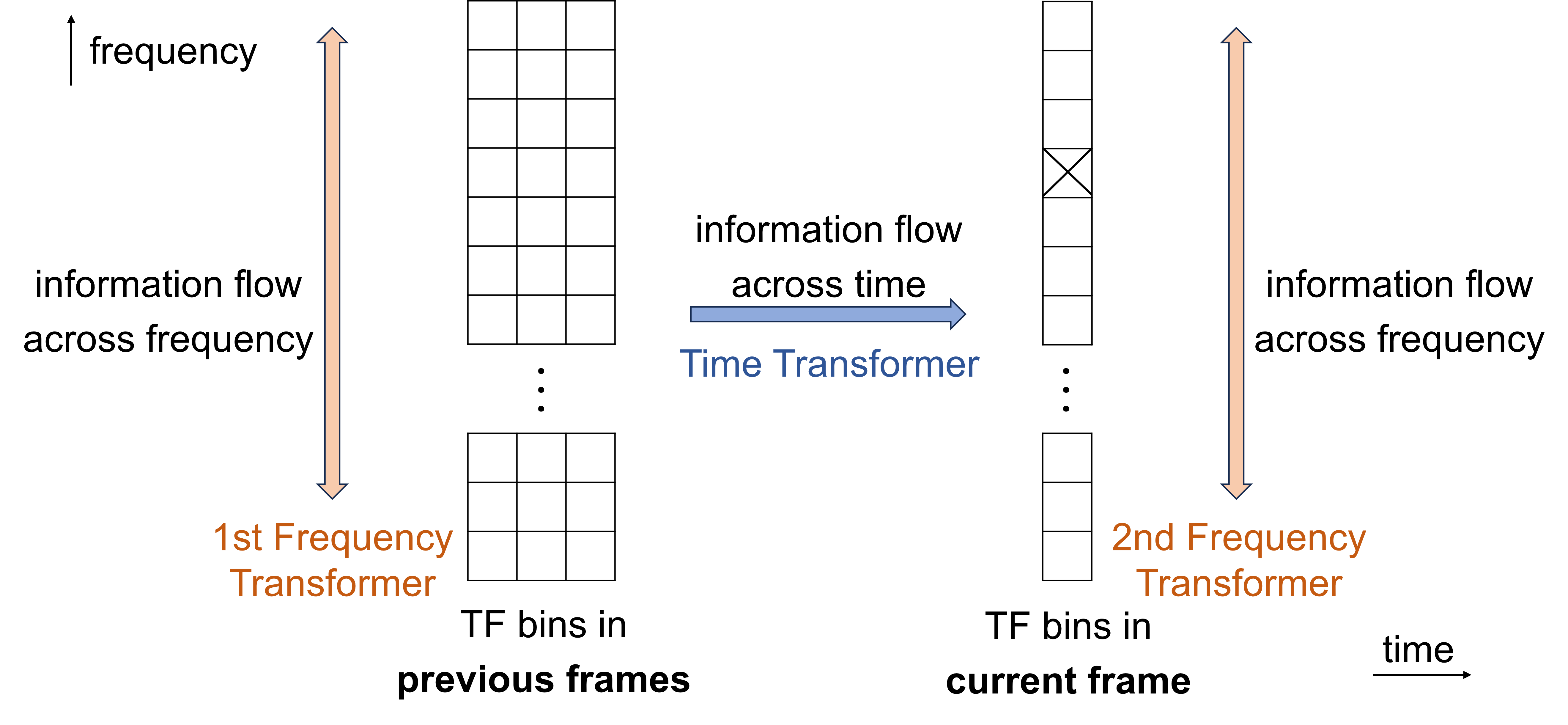}}
\caption{The schematic diagram for the information-exploitation flow of proposed efficient FTF-transformer structure. TF bins are represented by small blocks. Information flows from TF bins of previous and current frames to each certain TF bin (marked by a cross) are denoted by coloured arrows. The x and y axis are time and frequency dimensions, respectively.}  
\label{fig2}
\vspace{-0.4cm}
\end{figure}

The resulting `ladder-shaped' information-exploitation flow is presented in Fig.~\ref{fig2}. In the first frequency transformer, information is exchanged and explored along the frequency dimension, allowing each TF bin to gain context from other TF bins within the same frame. 
Subsequently, the causal time transformer propagates all dependencies from previous frames to the current frame. 
The second frequency transformer facilitates the re-exchange of updated information along the frequency dimension, thereby achieving global dependency exploitation.

The feature-dimension size quadratically contributes to the GRU complexity. Therefore, in the time transformer, instead of modelling the frequency dimension together with the feature dimension as in conventional GRUs, shared parameters of the GRU are utilised across frequency, for the temporal modelling, as shown in Fig.~\ref{fig1}. The global spectral dependency is nevertheless extracted by the anterior and posterior frequency transformers, with shared parameters across all frames. Compared to the conventional GRU bottleneck of CRUSE in \cite{braun2021towards}, the proposed FTF transformers reduces
Multiply-Accumulate Operations (MACs) and parameters by $27\%$ and $98\%$, respectively.

\subsection{Network Parameters and Training Target}
Encoder and decoder layers in the LCT-GAN model use a kernel size of (2, 3) and a stride of (1, 2) for time and frequency. The output channels are $\{16, 32, 64\}$ for the encoder layers and $\{32, 16, 1\}$, respectively, for the decoder layers. Padding is applied as necessary to ensure the preservation of dimensions.
    The hidden size is configured to be the same as the channel number of input feature for each grouped GRU layer, with a grouping factor of 4 applied. For the MHA, 4 heads are used.

For the training target, we applied the loss function in \cite{braun2021towards} including Mean Squared Error (MSE) and phase-aware MSE and extended it to a multi-resolution version \cite{yamamoto2020parallel}. The compression factor is set to 0.3, aligned with that in the network.
The sizes of multi-resolution STFT are defined as $m \in \{320, 512, 768\}$, all with an overlap of 50$\%$. As the mask estimate is for an STFT size of 512, the weight $\lambda_{512}$ is set to 2, twice that of others. The loss function for the generator comprises $L_\mathrm{adv\_gen}$ and the $L_\mathrm{multi\_res}$, the components of which are the same as \cite{kong2020hifi}. We observed that time-domain loss components $L_\mathrm{adv\_gen}$ and $L_\mathrm{adv\_dis}$ generally exhibited larger values compared to spectrogram-domain loss components $L_\mathrm{multi\_res}$. Therefore, their weights are set to $1\times10^{-2}$, to balance their contribution.

\section{Experiments}
\subsection{Experiment Setup}
Experiments were carried out on two widely-used datasets: Voicebank+Demand \cite{valentini2016investigating} and DNS3 challenge \cite{reddy21_interspeech}. For the experiments on DNS3 dataset, the validation and the 140-hour training dataset are generated from DNS3 wide-band English clean and noise dataset, encompassing Signal-to-Noise Ratios (SNRs) in the range of -5 to 20 dB.
The sampling rate is 16 kHz. The STFT window length is configured to 512, with a 50$\%$ overlap. 
The generator adopts the AdamW optimizer with a learning rate of $5 \times 10^{-4}$ and a batch size of 8, whereas the architecturally complex discriminators utilize a significantly lower learning rate to mitigate training instability induced by adversarial dominance. Specifically, $1 \times 10^{-7}$ is for the DNS3 dataset and $1 \times 10^{-9}$ is for Voicebank+Demand dataset.
The $\beta$ coefficients of (0.9, 0.99) and (0.8, 0.99) are assigned to the generator and discriminators, respectively.
\subsection{Studies on Validating the Efficacy of FTF Structure and Magnitude Estimation Within the Proposed LCT Framework}
\begin{table}[th]
\begin{threeparttable}
\caption{Results on Temporal / Spectral modelling and models with complex-valued outputs}
\label{tab:table1}
\centering
\setlength{\tabcolsep}{0.53mm}{
\begin{tabular}{ |c|l|ccc|ccc|} \hline
\multicolumn{2}{|c|}{\multirow{2}{*}{\textbf{Model}}} & \multicolumn{3}{c|}{\textbf{Synthetic Dataset}}  &\multicolumn{3}{c|}{\textbf{Real Recordings}}\\ 
\cline{3-8}
\multicolumn{2}{|c|}{} & \textbf{PESQ}  & \textbf{STOI} &\textbf{SI-SDR}  & \textbf{BAK} & \textbf{SIG}  & \textbf{OVL}\\ \hline
\multicolumn{2}{|c|}{Noisy speech} & 1.53 & 0.863 & 8.72 & 2.41 & 3.10 & 2.17 	  \\ \hline
\multirow{6}{*}{\textbf{Real-valued}} & TT                      & 2.18 & 0.895 & 14.01 & 3.67 & 2.93 & 2.54	  \\
\multirow{6}{*}{\textbf{Mask}}& FF                      & 2.34 & 0.908 & 14.24 & 3.98 & 2.69 & 2.45	  \\
\multirow{6}{*}{\textbf{Outputs}}& TF                      & 2.55 & 0.923 & 15.55 & \underline{4.00} & 2.89 & 2.64	  \\
& FT                      & 2.57 & 0.924 & 15.73 & 3.99 & 2.93 & 2.66	  \\
& TFT                     & 2.62 & 0.929 & 16.00 & 3.98 & 2.95 & 2.68	  \\
& \textbf{FTF (LCT)}      & \underline{2.68} & 0.932 & \underline{16.29} & \textbf{4.01} & 2.93 & 2.67	  \\
& TFTF                    & \underline{2.68} & \textbf{0.934} & \textbf{16.40} & 3.99 & \underline{2.97} & \underline{2.71}	  \\
& FTFT                    & \textbf{2.69} & \underline{0.933} & 16.25 & \underline{4.00} & \textbf{2.99} & \textbf{2.73}	  \\
  \hline
  
\multirow{2}{*}{\textbf{Complex-}} & LCT RI mapping      & 2.60 & 0.927 & 15.30 & 4.00 & 2.94 & 2.67	  \\
\multirow{2}{*}{\textbf{Valued}}& LCT RI masking      & 2.65 & 0.930 & 15.93 & \textbf{4.01} & 2.94 & 2.68	  \\
\multirow{2}{*}{\textbf{Outputs}}& LCT MCS mapping     & 2.63 & 0.928 & 15.20 & \textbf{4.01} & 2.95 & 2.69      \\
& LCT MCS hybrid      & 2.62 & 0.929 & 15.53 & \textbf{4.01} & 2.95 & 2.69	  \\
  \hline
	\end{tabular}}
 \begin{tablenotes}
\footnotesize
\item[a] All models are based on the generator network without discriminators.
\item[b] The T and F transformers used in this study have comparable MACs and parameters, thereby ensuring a fair basis for evaluation.
\item[c] All results are obtained causally without the need of look-ahead frames.
\end{tablenotes}
\end{threeparttable}
 \vspace{-0.4cm}
\end{table} 

Table~\ref{tab:table1} presents the ablation study on the bottlenecks formed by various time and frequency transformer combinations. Evaluations are on both the 1-hour unseen test set synthesised by DNS3 datasets and the official DNS3 real-recorded blind English test set. 
Compared to TT and FF models, TF and FT models demonstrate the efficacy of interleaved temporal and spectral modelling across metrics.
The comparison of TT and FF models highlights that spectral modelling achieves superior performance in wide-band Perceptual Evaluation of Speech Quality (PESQ) \cite{itut07}, Short-Term Objective Intelligibility (STOI) \cite{taal2010short}, Scale-Invariant Signal-to-Distortion Ratio (SI-SDR) \cite{le2019sdr}, and the BAK of DNS-MOS P.835 metrics \cite{reddy2022dnsmos}, while the temporal modelling enhances SIG more. 
These observations highlight the effectiveness of spectral modelling on noise suppression and overall quality enhancement.
Results on intrusive metrics indicate that the FTF structure offers the best trade-off with improvements saturating with more stages. As intrusive metrics reflect improvement with respect to ground truth, we select the FTF structure as the bottleneck of the proposed LCT network.


LCT models utilising the FTF bottleneck with complex-valued outputs such as Real-Imaginary (RI) masking and mapping, as well as Magnitude-Cos-Sin (MCS) mapping and hybrid \cite{10627167}, incur slightly more MACs. However, as demonstrated in Table~\ref{tab:table1}, they fail to improve performance in the intrusive metric scores. Thus, the complex-valued estimation is unnecessary in this context.

\subsection{Evaluations Against Baselines}
Next, we benchmark the proposed LCT-GAN model against SotA lightweight models \cite{schroeter2022deepfilternet, schroeter2022deepfilternet2, schroeter2023deepfilternet3, dang2023first} as well as contemporary, more sophisticated models \cite{zhao2022frcrn, shin22c_interspeech}. 
To ensure a fair comparison with SotA baselines ~\cite{schroeter2022deepfilternet2,schroeter2023deepfilternet3}, which employ a two-frame look-ahead yielding an overall latency of 40 $ms$, we adopt a one-frame look-ahead, resulting in a comparable latency of 48 $ms$. Additionally, experiments are conducted on the causal implementation of LCT-GAN with a 32 $ms$ latency.

\begin{table}[th]
\begin{threeparttable}
\caption{Evaluation results on Voicebank+Demand test dataset}
\label{tab:table2}
\centering
\setlength{\tabcolsep}{0.18mm}{
\begin{tabular}{ |l|c|c|c|c|c|c|c|c|} \hline
\multirow{2}{*}{\textbf{Model}} & \textbf{Param} & \textbf{MAC} & \multirow{2}{*}{\textbf{PESQ}}  & \multirow{2}{*}{\textbf{CSIG}} &\multirow{2}{*}{\textbf{CBAK}}  & \multirow{2}{*}{\textbf{COVL}} & \multirow{2}{*}{\textbf{STOI}}  & \multirow{2}{*}{\textbf{SI-SDR}}\\ 
& \textbf{[M]} & \textbf{[G/s]} &  &  &  &  &  & \\ \hline
{Noisy speech} & - & - & 1.97 & 3.34 & 2.44 & 2.63 & 0.92 & 8.4 	  \\ \hline
{FRCRN \cite{zhao2022frcrn}}                         & 10.27& 12.3 & \textbf{3.21} & 4.23 & 3.64 & 3.73 & - & -	  \\
{S-8.1GF \cite{shin22c_interspeech}}                & 1.38 & 4.07 & 3.12 & \textbf{4.45} & 3.61 & 3.82 & \textbf{0.95} & - \\ \hline
{CCFNet+$\triangle$ \cite{dang2023first}}                       & 0.62 & 1.47 & 3.03 & 4.27 & 3.55 & 3.61 & \textbf{0.95} & \textbf{19.1}	  \\ 
{CCFNet+Lite$\triangle$ \cite{dang2023first}}                 & 0.16 & 0.39 & 2.94 & -    & -    & -    & -     & 19.0	  \\
{DeepFilterNet \cite{schroeter2022deepfilternet}}     & 1.78 & \textbf{0.35} & 2.81 & 4.14 & 3.31 & 3.46 & 0.94 & 16.6 	  \\ 
{DeepFilterNet2 \cite{schroeter2022deepfilternet2}}   & 2.31 & 0.36 & 3.08 & 4.30 & 3.40 & 3.70 & 0.94 & - 	  \\ 
{DeepFilterNet3 \cite{schroeter2023deepfilternet3}}    & 2.31$\diamond$    & $0.36 \diamond$   & 3.17 & 4.34 & 3.61 & 3.77 & 0.94 & - 	  \\ 
 \hline
{LCT (Generator)}         & \textbf{0.14} & \textbf{0.35} & 3.01 & 4.38 & 3.64 & 3.75 & \textbf{0.95}  & 18.8	  \\ 
{LCT-GAN}     & \textbf{0.14} & \textbf{0.35} & 3.06 & \textbf{4.41} & \textbf{3.66} & 3.80 & \textbf{0.95}  & 18.8	  \\  
{LCT-GAN+PCS} & \textbf{0.14} & \textbf{0.35} & \textbf{3.20} & 4.39 & 3.46 & \textbf{3.84} & 0.94 &*	  \\ 
{LCT-GAN+PCS$\triangle$} & \textbf{0.14} & \textbf{0.35} & 3.16 & 4.35 & 3.45 & 3.81 & 0.94 & *	  \\ 
  \hline
	\end{tabular}}
 \begin{tablenotes}
\footnotesize
\item[a] Results of baselines are sourced from the respective references. A `-' indicates that the result is not reported in the corresponding reference. 
\item[b] '$\triangle$' indicates that the model is implemented in a causal manner.
\item[c] `*' denotes that SI-SDR is inapplicable for PCS training due to modified training targets.
\item[d] $\diamond$: MACs and footprint were not explicitly reported in \cite{schroeter2023deepfilternet3} but are expected to be similar to those in \cite{schroeter2022deepfilternet2}, given the minor modifications in \cite{schroeter2023deepfilternet3}.
\end{tablenotes}
\end{threeparttable}
 \vspace{-0.4cm}
\end{table} 

\textbf{Voicebank+Demand Dataset}:
In Table~\ref{tab:table2}, the evaluation results of the proposed LCT-GAN model on the Voicebank+Demand Dataset are presented in comparison to other baselines, along with their complexity and footprint, expressed by MAC/s and parameter numbers, respectively.
Composite instrumental metrics (CSIG, CBAK, COVL) \cite{hu2007evaluation}, are additionally introduced.
Utilising only $\approx\boldsymbol{6\%}$ of the parameters of DeepFilterNet2 and slightly fewer MACs, LCT-GAN achieves comparable, if not better scores across all metrics.
LCT-GAN does, however, exhibit a PESQ reduction of 0.11 compared to DeepFilterNet3.
Notably, when augmented with Perceptual Contrast Stretching (PCS) \cite{chao22_interspeech}—a technique designed to exaggerate speech features during training to improve intrusive metrics—LCT-GAN attains PESQ scores comparable to those of DeepFilterNet3 and generally surpasses all other lightweight models in all metrics.
In contrast, the causal CCFNet+(Lite), which entails slightly higher complexity and a larger footprint than LCT-GAN, demonstrates a marked PESQ degradation of 0.17 relative to the {\em causal} implementation of LCT-GAN with PCS.
Interestingly, our model variants are capable of surpassing even popular, more sophisticated models such as FRCRN \cite{zhao2022frcrn} and MANNER-S-8.1GF \cite{shin22c_interspeech}, which possess roughly one or two orders of magnitude larger MACs and parameters.

\begin{table}[th]
\vspace{-0.2cm}
\begin{threeparttable}
\caption{DNSMOS P.835 Evaluation results on DNS3 blind English test dataset -- mean and standard deviation of scores.}
\label{tab:table3}
\centering
\setlength{\tabcolsep}{2.5mm}{
\begin{tabular}{ |lccc|} \hline
{\textbf{Model}} 
 & \textbf{BAKMOS}  & \textbf{SIGMOS}   & \textbf{OVLMOS} \\ \hline
{Noisy speech}                                      & 2.41 ($\pm0.65$)   & 3.10 ($\pm0.47$) & 2.17 ($\pm0.43$) \\ 
{DeepFilterNet \cite{schroeter2022deepfilternet}}         & 3.83 ($\pm0.27$)& 3.02 ($\pm0.41$) & 2.69 ($\pm0.40$)\\
{DeepFilterNet2 \cite{schroeter2022deepfilternet2}}       & 3.95 ($\pm0.19$)  & \textbf{3.08} \textbf{($\pm$0.38)} & \underline{2.78} ($\pm$0.38)\\
{DeepFilterNet3 \cite{schroeter2023deepfilternet3}}       & 3.93 ($\pm0.20$)   & 3.02 ($\pm0.40$) & 2.72 ($\pm0.39$)	 \\ \hline
{LCT (Generator)}          & \underline{4.02} ($\pm$0.11) & 3.00 ($\pm0.40$) & 2.74 ($\pm0.40$) \\ 
{LCT-GAN}                                       & \textbf{4.04} \textbf{($\pm$0.09)} & \underline{3.06} ($\pm0.39$) & \textbf{2.80} \textbf{($\pm$0.38)} \\ 

  \hline
	\end{tabular}}
 \begin{tablenotes}
\footnotesize
\item[a] Results of baselines are obtained by official Python package v0.5.6 \cite{schroeter2022deepfilternet2}.
\item[b] Parameters and MACs of each model are the same as Table~\ref{tab:table2}.
\end{tablenotes}
\end{threeparttable}
\end{table} 

\textbf{DNS3 dataset}:
Considering the limited size of Voicebank+Demand, LCT-GAN is further trained on the synthesised DNS3 dataset and evaluated on the DNS3 real-recorded blind English test dataset \cite{reddy21_interspeech}. 
Due to the real-recorded nature of the test set, non-referential DNS-MOS metrics \cite{reddy2022dnsmos} are used. The results in Table ~\ref{tab:table3} indicate that the LCT-GAN outperforms all DeepFilterNet models in BAK and achieves considerable reduction in variance, suggesting better and more {\em consistent} noise suppression. 
Moreover, LCT-GAN remains consistent with SotA performance in DNS-MOS metrics.

The evaluation results of Table~\ref{tab:table2} and Table~\ref{tab:table3} reveal that the LCT model, even without discriminators and acting as a predictive model, can still achieve comparable performance to SotA models. This showcases the efficient latent-domain information exploitation of the FTF transformer in LCT. The ablation study of discriminators demonstrates that they can further contribute to improvements on speech quality metrics, especially in PESQ, DNSMOS- SIG \& OVL. This highlights that discriminators promote preserving speech components {\em while maintaining robust noise suppression} (no degradation on BAK). 

\textbf{Audio samples} showcasing the discussed result trends are available at \href{https://aspire.ugent.be/demos/EUSIPCO2025HZ/}{https://aspire.ugent.be/demos/EUSIPCO2025HZ/}. 
These samples further showcase the robust noise suppression and effective artefact elimination capacity of LCT-GAN. Notably, the spectral banding artefacts of DeepFilterNet are absent. 

\section{Conclusion}

We proposed a lightweight, causal, transformer-based GAN model designed for speech enhancement in resource-constrained applications.
The proposed FTF transformer block achieves an optimal trade-off between performance and computational efficiency by effectively capturing global dependencies through a refined interleaving of temporal and spectral modelling. This minimises both complexity and parameter footprint.
Experimental results indicate that the proposed LCT-GAN model yields SoTA performance in metrics among contemporary lightweight enhancement models, with considerable savings in parameters and MACs relative to these baselines. 
The additional ablation study on utilising discriminators demonstrates that multi-scale and multi-period discriminators result in improved speech component preservation. With PCS, a metric-enhancement strategy, the LCT-GAN can even exhibit performance that rivals more complex, representative models. However, some over suppression of weak fricatives and sibilants indicate areas for improvement in future work.

\bibliographystyle{IEEEtran}
\bibliography{mybib}

\end{document}